\documentclass{appolb}
\usepackage{epsfig}

\begin{document}
\title{INFORMATION THEORY POINT OF VIEW ON STOCHASTIC NETWORKS
\thanks{Presented at First Polish Symposium on Econo- and Sociophysics,
Warsaw, Poland, $19-20$ November  $2004$.}
}
\author{G.Wilk\thanks{e-mail: wilk@fuw.edu.pl}
\address{The Andrzej So\l tan Institute of Nuclear Studies,
         Ho\.za 69; 00-689 Warsaw, Poland}
\and  Z. W\l odarczyk\thanks{wlod@pu.kielce.pl}
\address{Institute of Physics, \'Swi\c{e}tokrzyska Academy,
         \'Swi\c{e}tokrzyska 15; 25-406 Kielce, Poland\\
         and\\
         University of Arts and Sciences (WSU), Weso\l a 52;
         25-353 Kielce, Poland
}
}
\maketitle
\begin{abstract}
Stochastic networks represent very important subject of research because they have
been found in almost all branches of modern science, including also sociology and
economy. We provide a information theory point of view, mostly based on its
nonextensive version, on their most characteristic properties illustrating it with
some examples.
\end{abstract}
 \noindent

\section{Introduction}
Different kinds of stochastic networks show up in nature whenever one is dealing with
complex systems of any kind \footnote{We refer to \cite{WW-net} for list of relevant
references to which we would like to add recent application of networks to describe
some features of multiparticle production processes in high energy hadronic
collisions \cite{WW-had}.}. There are two basic types of stochastic
networks:\\
$(a)$~~~Networks with constant number of nods, $M$, for which probability that given
node has $k$ connections with other nodes ($k$ links) is poissonian \cite{ER},
\begin{equation}
P(k) = \frac{\kappa_0^k}{k!}\cdot e^{-\kappa_0};\qquad \kappa_0 = \langle k\rangle .
\label{eq:ER}
\end{equation}
$(b)$~~~Networks in which number of nodes is not stationary and distribution of links
$P(k)$ is given by dynamics of the growth of network \cite{BAJ} and varies between
being {\it exponential},
\begin{equation}
P(k) = \frac{1}{m}\cdot \exp\left( - \frac{k}{m}\right), \label{eq:equal}
\end{equation}
(for the case when each new node connects with the already existing ones with equal
probability $\Pi(k_i) =1/(m_0+t-1)$ independent on $k_i$) or being {\it power-like},
\begin{equation}
P(k) = \frac{2m^2t}{m_0+t}\cdot k^{-3}, \label{eq:prefer}
\end{equation}
for the case of preferential attachment (the so called "rich-get-richer" mechanism,
here $m<m_0$ is the number of new nodes added in each time step) with, in this case,
$\Pi(k_i)=k_i/(2mt)$ choice.\\

Recently we have demonstrated that using information theory approach in its
nonextensive version (i.e., maximizing Tsallis entropy \cite{T}
$H_q=[1-\sum_kP^q_q(k)]/(q-1)$ under conditions that $\langle k\rangle_q=\sum_k k
P^q_q(k)/\sum_k P^q_q(k) = \kappa_0$ and $\sum_k P_q(k) =1$) one obtains that
\cite{WW-net}
\begin{equation}
P_q(k) =\frac{1}{\kappa_0}\left[1 - (1-q)\frac{k}{\kappa_0}\right]^{\frac{q}{1-q}},
\label{eq:Pq}
\end{equation}
which is very universal because for $q\rightarrow 1$ it recovers eq. (\ref{eq:equal})
whereas for $k>>\kappa_0/(q-1)$ it leads to the power-like distribution $P_q(k)
\propto k^{q/(1-q)}$. For properly chosen parameters $\kappa_0$ and $q$ it is
therefore able to describe data in the {\it whole region} of variable $k$. As
example, in \cite{WW-net} distribution of WWW network after \cite{JK} with the mean
number of connections $\langle k\rangle =5.46$ has been described by $P_q(k)$ with
parameters $\kappa_0=1.91$ and $q=1.65$ obtaining for large values of $k$ power-like
distribution $\propto k^{-\gamma}$ with $\gamma = q/(q-1)=2.54$ as observed in
\cite{JK}. Notice that condition of finiteness of the first moment of $P_q(k)$, equal
to $\langle k\rangle =\kappa_0/(2-q)$, results in the limitation that $q<2$, whereas
similar condition imposed on the variance of $P_q(k)$, given by $Var(k) =
\kappa_0^2/[(3-2q)(2-q)^2]$, limits $q$ further to $q<3/2$ (it is worth to stress
that for $q=3/2$, at which variance $Var(k)$ diverges, one gets exponent $\gamma =
3$, as in eq. (\ref{eq:prefer})).\\

Since then detailed analysis of preferential attachment growth and its connection
with nonextensive statistical mechanics has been performed in \cite{T-net} whereas in
\cite{Abe-net} it was demonstrated that introducing notion of fluctuations to a
random graph one obtains, as one of the results, the scale-free power-like networks
(in very much similar way as fluctuations of parameter $1/m$ in distribution
(\ref{eq:equal}) given entirely by parameter $q$ lead directly to distribution
$P_q(k)$ \cite{WW}). In what follows we shall provide two additional particular
examples of using methods described in \cite{WW-net,WW} to analysis of some
stochastic network features.\\

\section{Transport on network}

Investigations of transport in the internet \cite{TR,AS} show that distribution of
travel times can be described by power distribution of the form $P(t)\propto
t^{-(\alpha+1)}$. Such form implies specific dynamic of transport on the network in
which many correlated packets travels at the same time. Detailed description of such
dynamical picture is particularly complicated because of the lack of some global
navigation prescription, by the fact that packets can be send parallel fashion and,
finally, by dependence on the structure of network and on the algorithm of selecting
the transportation path used.\\

To describe this complicated character of transport let us introduce transport rate
distribution function $f(\tau)$ and write distribution of the transportation times as
\begin{equation}
P(t) = \int_0^{\infty}d\tau f(\tau) \exp\left( - \frac{t}{\tau}\right) .
\label{eq:ftau}
\end{equation}
Function $f(\tau)$ can be obtained by considering stochastic process given by the
following Langevin equation
\begin{equation}
\frac{d\tau}{dt} + [1 + \xi(t)]\cdot \tau = \tau_0 , \label{eq:langtau}
\end{equation}
where $\xi (t)$ is white gaussian noise with $\langle \xi(t)\rangle =0$ and $\langle
\xi(t)\xi(t+\Delta t)\rangle = 2D\delta(\Delta t)$, parameter $\tau_0$ is the
characteristic transport time on network and $D$ denotes variation of transport
times. Considering now the corresponding Fokker-Planck equation one gest (cf.
\cite{WW}) that\footnote{It should be noticed that distribution just obtained is the
most expected one from the point of view of the maximization of Shannon information
entropy, $H=-\sigma_k P(k)\ln P(k)$ under constraints that $\int f(\tau)d\tau =1$,
$\langle \tau\rangle = \tau_0$ and (because distribution we are looking for is
defined only for $\tau >0$) that $\langle \ln (\tau)\rangle = \ln(\tau_0/D)$
\cite{MS}.}
\begin{equation}
f(\tau) = \frac{\mu^{\nu}}{\Gamma(\nu) \tau^{\nu -1}} \exp\left( -
\frac{\mu}{\tau}\right), \label{eq:FPtau}
\end{equation}
where $\mu = \nu \tau_0$ and $\nu = 1/D$., i.e., it is given by gamma distribution in
the variable $1/\tau$. As was shown in \cite{WW} such function $f(\tau)$ results in
the power-like distribution of $P(t)$,
\begin{equation}
P(t) = \frac{2-q'}{\tau_0}\left[1 - (1 -
q')\frac{t}{\tau_0}\right]^{\frac{1}{1-q'}},\qquad {\rm where}\qquad q'=1+D .
\label{eq:Ptransp}
\end{equation}
In \cite{AS} it has been shown that exactly such form is observed experimentally. It
means therefore that approach based on Tsallis statistics describes stationary states
on Internet.\\

Power-like asymptotic form of $P(t)\propto t^{-(\alpha +1)}$ implies existence of
long-range correlations in the network communication with formation of large active
aggregates of transportation streams \cite{HR}. Denoting aggregates by the number of
active streams existing in time $t$, $K_i$, these long-range correlations are given
by autocovariance $R(t)=cov(K_t,K_{t+s})$ (typically, for the internet transport, the
written above power-like form of $P(t)$ leads directly to similar power-like form of
$R(t) \propto t^{-(\alpha-1)}$ with $\alpha \in (1,2)$),
\begin{equation}
R(t) = \int_t^{\infty}dy [1 - F(y)], \label{eq:R}
\end{equation}
where $F(t)$ denotes distribuant of $P(t)$ \cite{RS}. It turns out that using
formalism of random matrices (where matrix elements $W_{ij}$ have random probability
distributions) and identifying the number of decay channels with the size $K$ of the
aggregate one can immediately obtain that \cite{rWW} exponent $\alpha$ above is given
by the mean size of the aggregate $\langle K\rangle$:
\begin{equation}
\alpha = \frac{2-q'}{q'-1} = \frac{1}{D} - 1 = \frac{1}{2}\langle K\rangle - 1 .
\label{eq:aggregate}
\end{equation}
Notice that with increasing size of the aggregate, $\langle K\rangle$, fluctuations
in $f(\tau)$ diminish, $\langle(\tau^{-1}-\langle\tau^{-1}\rangle)^2\rangle =
2\langle \tau^{-1}\rangle^2/\langle K\rangle$, and in the limit of large $\langle
K\rangle$ the power-like behavior in (\ref{eq:Ptransp}) becomes exponential one (as
now $q'\rightarrow 1$.)

\section{Epidemic dynamics in network}

let us now concentrate on the dynamic of spreading of viruses using simple model of
the SIS type (susceptible-infects-susceptible). In this model each node can be either
"healthy" ($H$) or "infected" ($I$) and infection spread up due to connections
between nodes. On every time step node $H$ is infected with probability $\nu$ if it
is linked with at least one link with node $I$. At the same time node $I$ is cured
with probability $\delta$ (i.e., the mean time of infection is equal $D=1/\delta$).
Denoting the effective rate of diffusive spreading up the virus by $\lambda =
\nu/\delta$ it can be shown \cite{DB} that probability density of nodes $I$ is given
by
\begin{eqnarray}
\rho = 0\qquad &{\rm for}&\qquad \lambda < \lambda_c; \nonumber\\
\rho \propto \left(\lambda - \lambda_0\right)^{\beta} &{\rm for}&\qquad \lambda
>\lambda_c ,\label{eq:diff}
\end{eqnarray}
i.e., virus which spreading rate $\lambda$ exceeds the threshold value $\lambda_c$
survives whereas it quickly vanishes when his spreading rate is below the epidemic
threshold \cite{PV,AJB}. This threshold depends on the variation of the number of
connections in the network, $\lambda_c = \langle k\rangle/\langle k^2\rangle$. In
regular networks (where $P(k)=\delta(k-k_0)$) and in stochastic networks (where
$P(k)$ is poissonian) $\langle k^2\rangle$ is always finite and therefore $\lambda_c$
is always greater than zero. Situation changes drastically for scale-free networks
for which $P(k) \propto k^{-\gamma}$, with $\gamma \leq 3$. In this case for $\gamma
\rightarrow 3$ one has $\lambda_c\rightarrow 0$ and there is {\it no threshold for
the epidemic}\footnote{For example, for $P(k)$ given by eq. (\ref{eq:Pq}) $\lambda_c
= (3-2q)/(2\kappa_0)$, which for $q\rightarrow 3/2$ tends to
zero.}.\\

Let us analyze this in more detail. Denoting by $x_i$ the part of nodes of type $i$
(i.e., with $i$ connections) which are able to be infected and by $y_i$ the part of
already infected nodes, one can write following evolution equation (where $\nu_{ij} =
ij\nu/\langle k\rangle$):
\begin{equation}
\frac{\partial x_i}{\partial t} = - x_i\, \sum_i \nu_{ij} y_j; \qquad \qquad
\frac{\partial y_i}{\partial t} = x_i\, \sum_i \nu_{ij} y_j - y_i \delta .
\label{eq:partial}
\end{equation}
Solving (\ref{eq:partial}) one can shown \cite{AM} that the part of nodes which are
{\it always} infected (it is called {\it the final epidemic size}) is given by
\begin{equation}
I = \left\langle 1 - \exp( - k\alpha) \right\rangle\qquad {\rm where}\qquad \alpha =
\rho_0\frac{\langle k - k\exp(-k\alpha)\rangle}{\langle k^2\rangle} .\label{eq:I}
\end{equation}
Here $\rho_0=\nu D\langle k\rangle$ is the mean number of the secondary infections
caused by introducing one infected node in the network (under assumption that each
node has exactly $\langle k\rangle$ connections).\\

For regular network with $P(k)$ given by Poisson distribution (eq. (\ref{eq:ER})) the
size of epidemy is
\begin{equation}
I = 1 - \exp\left[ - \langle k\rangle\left( 1 - e^{-\alpha}\right)\right]
\label{eq:sizeER}
\end{equation}
where $\alpha$ is given by the following transcedental equation:
\begin{equation}
\alpha = \frac{\rho_0}{\langle k\rangle}\left\{1 - \exp\left[- \alpha - \langle
k\rangle\left(1 - e^{-\alpha}\right)\right]\right\} . \label{eq:alpha}
\end{equation}
Assuming small $\alpha$ and keeping in (\ref{eq:alpha}) only linear terms in $\alpha$
one obtains that
\begin{equation}
\alpha \simeq \frac{2\left[ 1 + \langle k\rangle \left(1 -
\frac{1}{\rho_0}\right)\right]}{(1+\langle k\rangle)^2 + \langle k\rangle} ,
\label{eq:aaa}
\end{equation}
from which one can estimate the epidemic threshold (which is for $I=0$, i.e., also
for $\alpha =0$) as being given by
\begin{equation}
\frac{1}{\rho_0} = 1 + \frac{1}{\langle k\rangle} . \label{eq:thr}
\end{equation}
\vspace{-7mm}
\begin{figure}[ht]
\begin{center}
        \epsfig{file=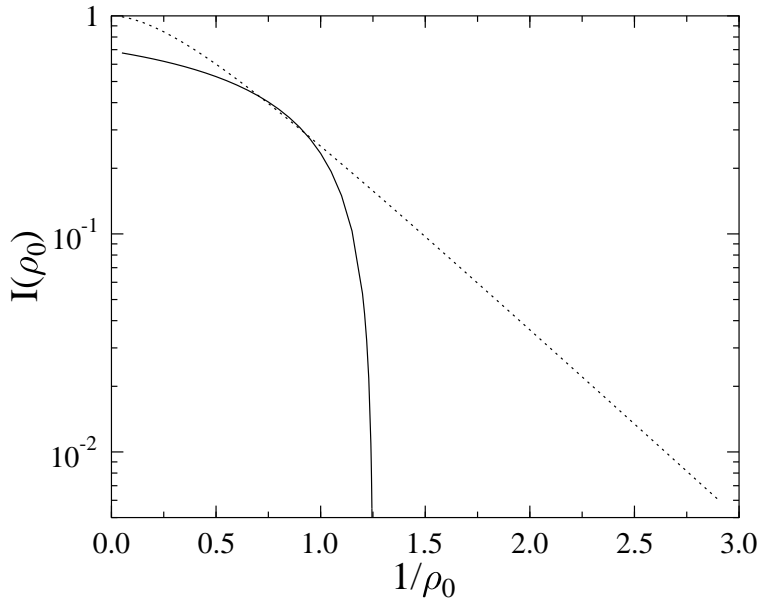, width=120mm}
\end{center}
  \caption{Examples of behavior of the final epidemic size $I(\rho_0)$
           as function of inverse of the mean number of secondary infections
           caused by introduction to network single infected virus, $\rho_0$,
           for network with mean number of connections equal $\langle k\rangle =4$.
           Full line corresponds to poissonian distribution of links (regular
           networks), dashed line is for power-like distributions of links
           (scale-free networks). Notice that in the later case there are
           situations in which there is no threshold to epidemic.}
  \label{Figure1}
\end{figure}

For the scale free network with $P(k)$ given by eq. (\ref{eq:prefer}), for small
values of $\rho_0$ we get the following size of epidemy:
\begin{equation}
I \sim 2e^{0.423}\cdot \exp\left( - \frac{2}{\rho_0}\right) . \label{eq:pk3}
\end{equation}
In the general case described by $P_q(k)$ as given by eq. (\ref{eq:Pq}) one has, in
the approximation that $\kappa_0\alpha >1$, that
\begin{equation}
I = 1 - \frac{1}{\alpha}\cdot \frac{\langle k\rangle}{q(2-q)}, \label{eq:qsize}
\end{equation}
where $\alpha$ is given by the equation
\begin{equation}
\alpha = \frac{\rho_0}{\langle k\rangle}\left[ 1 - \frac{1}{\alpha^2q(2-q)}\right]
.\label{eq:aq}
\end{equation}
As is clearly demonstrated in Fig. 1 only in the former case there is
epidemic threshold, in the later one no such effect appears.\\

\section{Conclusions}

Stochastic networks represent very complex phenomena of different origin. As such
they have been also investigated by using statistical mechanic approach \cite{MW}.
Here we have shown that they can be also described (at least in what concerns some of
their most commonly demonstrated properties) by using information theory approach
based on Tsallis statistics. Such approach allows to describe by means of single
formula probabilities of number of connections in a given network, $P_q(k)$, in
essentially all kinds of networks, from purely exponential ones (with $q=1$ to a
scale-free power-like ones with exponent $\gamma = q/(q-1)$. In this approach it is
clear that $\gamma=3$ found in many systems with complex topologies correspond to
$q=3/2$, a value for which variance of $P_q(k)$ diverges.\\

We have presented two examples of investigations of networks by means of Tsallis
power-like formula (\ref{eq:Pq}), transport on networks and development of epidemic
on networks. Both have numerous (and still growing) practical applications to which
others are added (like, for example, description of the earthquakes statistics
\cite{Aq}). As example we provide here two result, one concerning description of
known distributions of sexual partners, in Fig. 2, and one illustrating distribution
of populations of different agglomerations, in Fig. 3. Both can be very well
described by means of only two parameters, $\kappa_0$ and $q$. The later parameter
describes, according to \cite{WW-net,T-net,Abe-net}, in a summarily way the influence
dynamic of the network (resulting in intrinsic fluctuations, correlations, fractal
structure and the like). Let us close with remark that different behavior of
different epidemics (existence of threshold or not) is naturally explained by the
different character of networks of contacts causing infection: those corresponding to
regular networks (like, for example, flue epidemic mostly following earth
communication network) show some threshold whereas those corresponding with
scale-free networks (like, for example AIDS epidemic) have no
thresholds.\\

\vspace{-7mm}
\begin{figure}[ht]
\begin{center}
        \epsfig{file=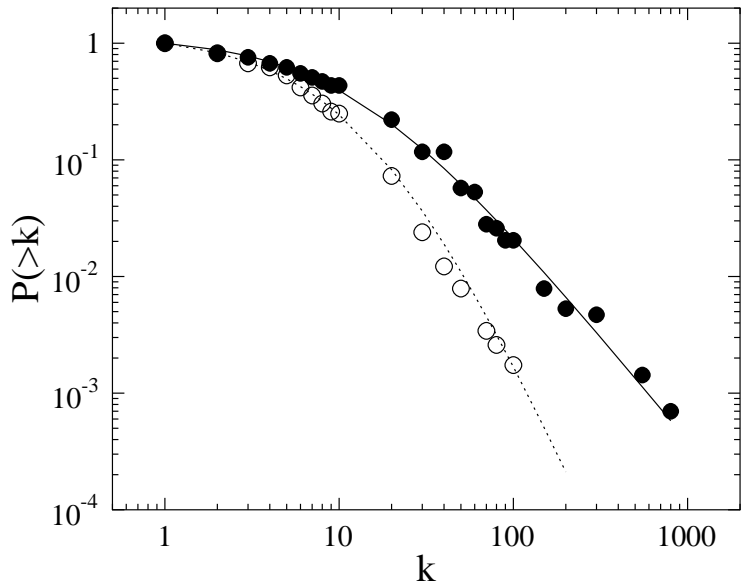, width=120mm}
\end{center}
\vspace{-5mm}
  \caption{Cumulative probability distribution of the number
           of sexual partners as given by \cite{Partners}
           compared with eq. (\ref{eq:Pq}) for men (full line and symbols;
           here $q=1.55$ and $\kappa_0 = 0.68$ resulting in
           $\langle k\rangle = 15$) and women (dashed line, open symbols;
           here $q=1.3$ and $\kappa_0 = 0.48$ resulting in
           $\langle k\rangle = 7$.}
  \label{Figure2}
\end{figure}

\vspace{-7mm}
\begin{figure}[ht]
\begin{center}
        \epsfig{file=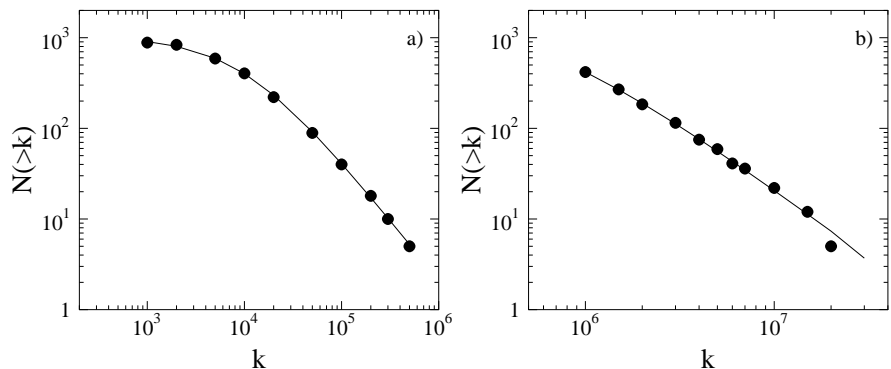, width=120mm}
\end{center}
\vspace{-5mm}
  \caption{Examples of unnormalized cumulative probability distributions
           of encountering cities in Poland
           with population greater than some given value $k$ on this value,
           $(a)$, and the same for agglomerations in the whole world, $(b)$.
           Data (points) were taken directly from \cite{Agglodata} whereas
           curves represent fits to Tsallis formula: $N(>k) = C\cdot P_q(q)$
           (with the following sets of parameters: $C=1032$, $\kappa_0 =7280$
           and $q=1.75$ for $(a)$ and $C= 2362$, $\kappa_0=310000$ and $q=1.65$
           for $(b)$). For other similar results see \cite{Cities}.}
  \label{Figure3}
\end{figure}


\newpage

\begin{thebibliography}{99}

\bibitem{WW-net} G.Wilk and Z.W\l odarczyk, {\sl Acta Phys. Polon.}
                 {\bf B35} (2004) 871.

\bibitem{WW-had} G.Wilk and Z.W\l odarczyk, {\sl Acta Phys. Polon.}
                 {\bf B35} (2004) 2143.

\bibitem{ER} P.Erd\"os and A.R\'enyi, {\sl Pub. Math. Inst. Hung.
             Acad. Sci.} {\bf 5} (1960) 17.

\bibitem{BAJ} A.L.Barabasi, R.Albert and H.Jeong, {\sl Physica} {\bf
              A272} (1999) 173.

\bibitem{T} C.Tsallis, {\sl J. Stat. Phys.} {\bf 52} (1988) 479;
            cf. also C.Tsallis, in {\it Nonextensive Statistical Mechanics
            and its Applications}, S.Abe and Y.Okamoto (Eds.), Lecture Notes
            in Physics LPN560, Springer (2000). For updated bibliography
            on this subject see http://tsallis.cat.cbpf.br/biblio.htm.

\bibitem{JK} H.Jeong and B.Kahn, {\it Complex Scale-Free Networks},
             in {\sl Asia Pacific Center for Theoretical Physics
             Bulletin} 07-08 (2001) 3.

\bibitem{T-net} D.J.B.Soares, C.Tsallis, A.M.Mariz and L.R.da Silva, {\it
                Preferential attachement growth model and nonextensive
                statistical mechanics}, cond-mat/0410459, to be published
                in {\sl Europh. Lett.} (2005); cf. also: C.Tsallis, M.Gell-Mann
                and Y.Sato, {\it Special scale-invariant occupancy of phase
                space makes the entropy $S_q$ additive}, cond-mat/0502274.

\bibitem{Abe-net} S.Abe and S.Thurner, {\it Analytic formula for hidden variable
                  distribution: complex networks arising from fluctuating
                  random graphs}, cond-mat/0501429.

\bibitem{WW} G.Wilk and Z.W\l odarczyk, {\sl Phys. Rev. Lett.}
             {\bf 84} (2000) 2770, {\sl Chaos, Solitons and
             Fractals} {\bf 13/3} (2001) 581 and {\sl Physica} {\bf
             A305} (2002) 227.

\bibitem{TR} B.Tadic and G.J.Rodgers, {\sl Complex Syst.} {\bf 5} (2002) 445.

\bibitem{AS} S.Abe and N.Suzuki, {\sl Phys. Rev.} {\bf E67} (2003) 016106.

\bibitem{MS} E.W.Monroll and M.F.Schlesinger, {\sl J.Stat. Phys.} {\bf 32} (1983)
             209.

\bibitem{HR} D.Heath and S.Resnik and G.Samorodnitsky, {\sl Math. Oper. Tres.}
             {\bf 23} (1998) 145.

\bibitem{RS} S.Resnick and G.Smorodnitsky, {\sl Queueing Syst.} {\bf 33} (1999) 43.

\bibitem{rWW} G.Wilk and Z.W\l odarczyk, {\sl Phys. Lett.} {\bf A290} (2001) 55.

\bibitem{DB} Z.Dezso and A.L.Barabasi, {\sl Phys. Rev.} {\bf E65} (2002)
             055103(R).

\bibitem{PV} R.Pastor-Satorras and A.Vespignani, {\sl Phys. Rev. Lett.}
             {\bf 86} (2001) 3200.

\bibitem{AJB} R.Albert, H.Jeong and A.L.Barabasi, {\sl Nature} {\bf 406}
              (2000) 378.

\bibitem{AM} R.M.Anderson and R.M.May, {\it Infectious Diseases of Humans: Dynamics
             and Control}, Oxford University Press, Oxford 1991.

\bibitem{Partners} F.Liljeros, C.R.Edling, L.A.N.Amaral, H.E.Stanley and Y.Aberg,
                   {\sl Nature} (2001) 907.

\bibitem{MW} Cf., for example, A.Majka and W.Wi\'slicki, {\sl Physica} {\bf A337}
             (2004) 645 and references therein.

\bibitem{Aq} S.Abe and N.Suzuki, {\sl Physica} {\bf A337} (2004) 357.

\bibitem{Agglodata} See data basis in http://www.citypopulation.de.

\bibitem{Cities} L.C.Malacarne, R.S.Mendes and E.K.Lenzi, {\sl Phys. Rev.} {\bf E65}
                 (2001) 017107.

\end{thebibliography}
\end{document}